\documentclass[conference]{IEEEtran}
\ifCLASSINFOpdf
\else
\fi
\usepackage{hyperref}
\usepackage{soul}
\newcommand{\cev}[1]{\reflectbox{\ensuremath{\vec{\reflectbox{\ensuremath{#1}}}}}}

\usepackage{amsmath}
\usepackage{authblk}
\usepackage{bm}
\usepackage{xcolor,colortbl}
\usepackage{stackengine}

\usepackage{cite}
\usepackage{amssymb}
\usepackage{mathtools}
\usepackage{eucal}

\usepackage{graphicx}
\usepackage{pgfplots}
\usepackage{tikz}
\usetikzlibrary{patterns}
\usepackage{array}
\usepackage{subfigure}

\graphicspath{ {./images/} }
\usepackage[utf8]{inputenc}
\usepackage{comment}
\usepackage[utf8]{inputenc}
\usepackage{multirow}
\usepackage{url}
\usepackage{multicol}
\usepackage{xcolor}
\usepackage[linesnumbered,ruled,vlined]{algorithm2e}
\usepackage{algorithmic}
\usepackage{xcolor}
\usepackage{soul}
\usepackage{float}

\SetCommentSty{mycommfont}

\pgfplotsset{compat=1.17} 
\begin{document}
\title{Real-Time Massive MIMO Channel Prediction: A Combination of Deep Learning and NeuralProphet}
\author[1, 2]{M. Karam Shehzad}
\author[1]{Luca Rose}
\author[3]{M. Furqan Azam}
\author[2]{Mohamad Assaad}
\affil[1]{Nokia Bell-Labs\\France.}
\affil[2]{Laboratoire des Signaux et Systèmes, CentraleSupelec, CNRS, University of Paris-Saclay, France.}
\affil[3]{Flemish Institute for Technological Research (VITO), 2400 Mol, Belgium.}
\affil[ ]{ \{muhammad.shehzad, luca.rose\}@nokia.com\\
muhammadfurqan.azam@vito.be, mohamad.assaad@centralesupelec.fr}
\maketitle

\begin{abstract}
Channel state information (CSI) is of pivotal importance as it enables wireless systems to adapt transmission parameters more accurately, thus improving the system's overall performance. However, it becomes challenging to acquire accurate CSI in a highly dynamic environment, mainly due to multi-path fading. Inaccurate CSI can deteriorate the performance, particularly of a massive multiple-input multiple-output (mMIMO) system. This paper adapts machine learning (ML) for CSI prediction. Specifically, we exploit time-series models of deep learning (DL) such as recurrent neural network (RNN) and Bidirectional long-short term memory (BiLSTM).
Further, we use NeuralProphet (NP), a recently introduced time-series model, composed of statistical components, e.g., auto-regression (AR) and Fourier terms, for CSI prediction. Inspired by statistical models, we also develop a novel hybrid framework comprising RNN and NP to achieve better prediction accuracy. The proposed channel predictors (CPs) performance is evaluated on a real-time dataset recorded at the Nokia Bell-Labs campus in Stuttgart, Germany. Numerical results show that DL brings performance gain when used with statistical models and showcases robustness.       
\end{abstract}
\IEEEpeerreviewmaketitle
\section{Introduction} \label{sec1}
Artificial intelligence (AI) and machine learning (ML) are the defining technologies of next-generation wireless networks, called sixth-generation (6G). It is expected that AI/ML will play a pivotal role in the design phase of 6G wireless networks \cite{Karam_AI_6G, AI_Jakob}. In this regard, standardization of ML has also begun \cite{Karam_AI_6G}. Specifically, it is expected that ML, together with massive multiple-input multiple-output (mMIMO), disruptive technology of fifth-generation (5G), can improve, for instance, precoding gain \cite{karam_RNN_ICC, Karam_AI_6G}. Furthermore, the validity of ML algorithms in a real-time environment has opened up a new horizon for the consideration of ML in 6G \cite{Karam_WCL}.

A mMIMO system can improve signal-to-noise ratio, as well as throughput, by utilizing diversity and multiplexing techniques, respectively \cite{mMIMO}. However, accurate channel state information (CSI) is indispensable to get expected gain. In a highly dynamic wireless communications environment, it is hard to acquire accurate CSI, e.g., due to reporting compressed CSI to base station (BS) by user-equipment (UE), and feedback/processing delays \cite{Karam_KF}.

To acquire accurate CSI, researchers have started exploiting an \textit{active} approach, known as channel prediction \cite{Karam_WCL} as it can improve the accuracy of CSI without requiring extra radio resources. The key idea of channel prediction is to forecast CSI realizations that can mitigate, e.g., compressed CSI and induced delays. Recently, its application to reduce mMIMO CSI feedback overhead and accuracy improvement of acquired CSI at BS have opened new doors for its consideration \cite{karam_RNN_ICC, Karam_KF}. 

The study of channel prediction has been considered by a few researchers in the literature \cite{AR_Model, Parametric_Model, Karam_WCL, W_Jiang_CP, recurrent}, which is mainly divided into statistical models and ML. For example, auto-regression (AR) and parametric models have been studied in \cite{AR_Model} and \cite{Parametric_Model}, respectively. However, the downside of statistical models is their iterative re-estimation of parameters that can expire quickly in a dynamic environment. 
And due to manipulation of matrices, parameters re-estimation can be costly \cite{Parametric_Model}. In contrast, ML has the capability of making multi-step prediction, as well as can provide huge gains in a diverse environment. To this end, \cite{W_Jiang_CP} and \cite{recurrent} evaluated the performance of a recurrent neural network (RNN), an ML algorithm, on a synthetic dataset. To demonstrate the effectiveness of RNN in a real-world environment, \cite{Karam_WCL} evaluated the performance of RNN on compressed and uncompressed CSI. 

In this paper, we evaluate the performance of various state-of-the-art deep learning (DL) models such as RNN \cite{Karam_WCL} and Bidirectional long-short term memory (BiLSTM). Also, a DL inspired statistical algorithm, i.e., NeuralProphet (NP), is tested on out-of-sample data. Further, getting the inspiration from using statistical algorithm and DL together, we propose a novel hybrid framework composed of RNN and NP, which yields better prediction results than individual models. 
In addition to this, we employ hyperparameter tuning for each of these individual models to select only the best training parameters. We observe the performance of channel predictors (CPs) in a realistic environment.

Rest of the paper is organized as follows: Section\,\ref{channel prediction} provides details of CPs used in our study. Section\,\ref{Measurement Campaign}  summarizes real-time dataset. Section\,\ref{results} gives performance comparisons of CPs, and Section\,\ref{conclusion} concludes the paper.   

\section{Channel Prediction}\label{channel prediction}
In this section, we highlight the channel prediction models that are used in this paper. They are divided into three parts: RNN, BiLSTM, and a hybrid model.

\subsection{Recurrent Neural Network (RNN)}\label{RNN}
RNN has emerged as a promising technique for time-series predictions \cite{RNN_Jordan}. 
Due to the presence of recurrent components, its prediction capability surpasses feed-forward neural networks (NNs). 
In this study, we use RNN to design a hybrid model. Given the vastity of the available literature on RNN \cite{Karam_WCL}, we do not detail its structure. The interested reader is referred to \cite{Karam_WCL} and \cite{recurrent} to get deeper understanding. In summary, RNNs are fed with $d$-step delayed inputs along with corresponding labels to generate multi-step ahead predicted CSI realizations. 
Later in Section\,\ref{Hybrid}, we will utilize the predicted channel vector of RNN for the design of the hybrid model. 
\begin{figure*}[ht]
    \centering
    \subfigure[Single-cell of LSTM.]{\includegraphics[width=0.49\textwidth]{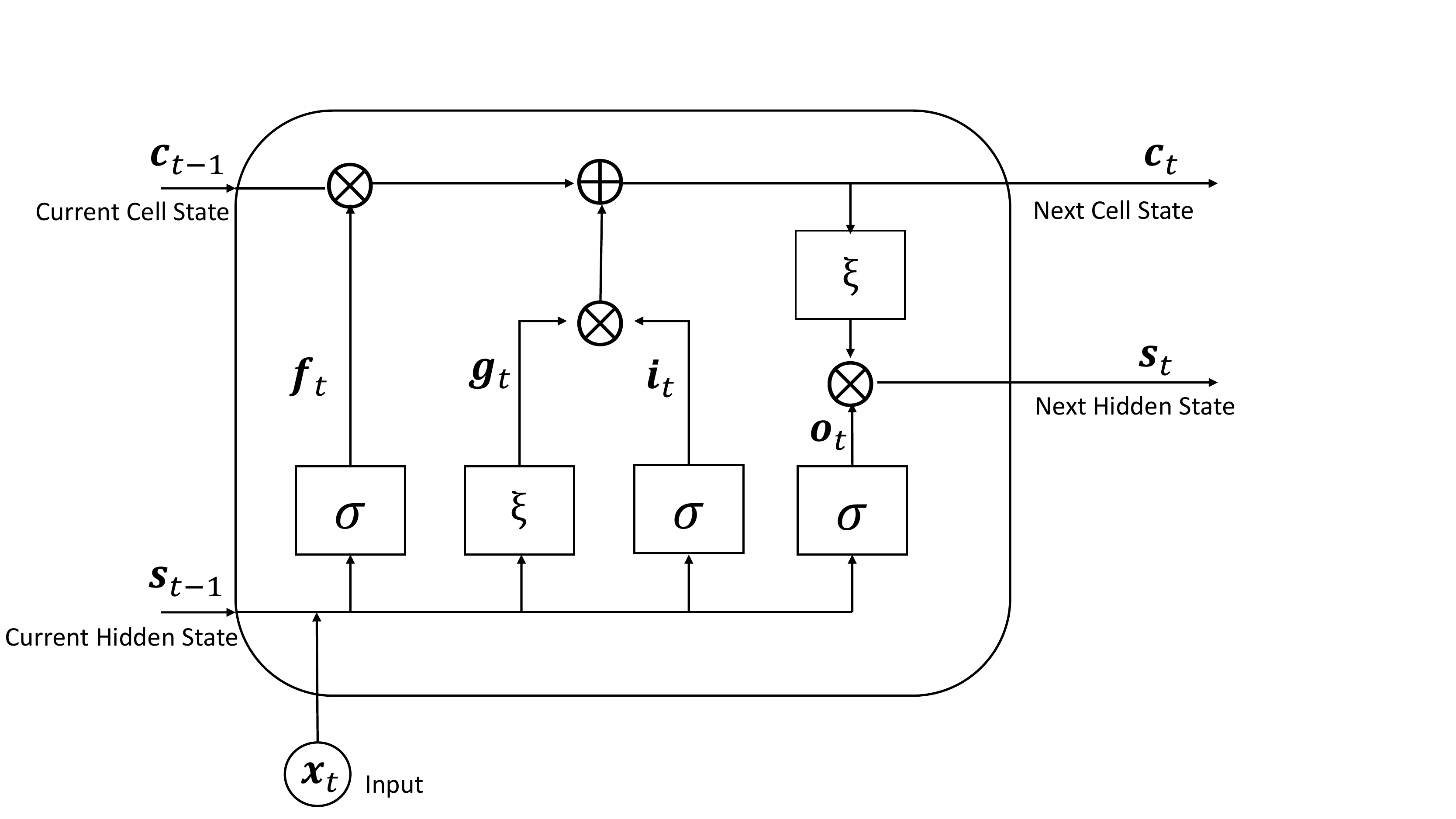}}
        \subfigure[Fully connected BiLSTM-based NN.]{\includegraphics[width=0.49\textwidth]{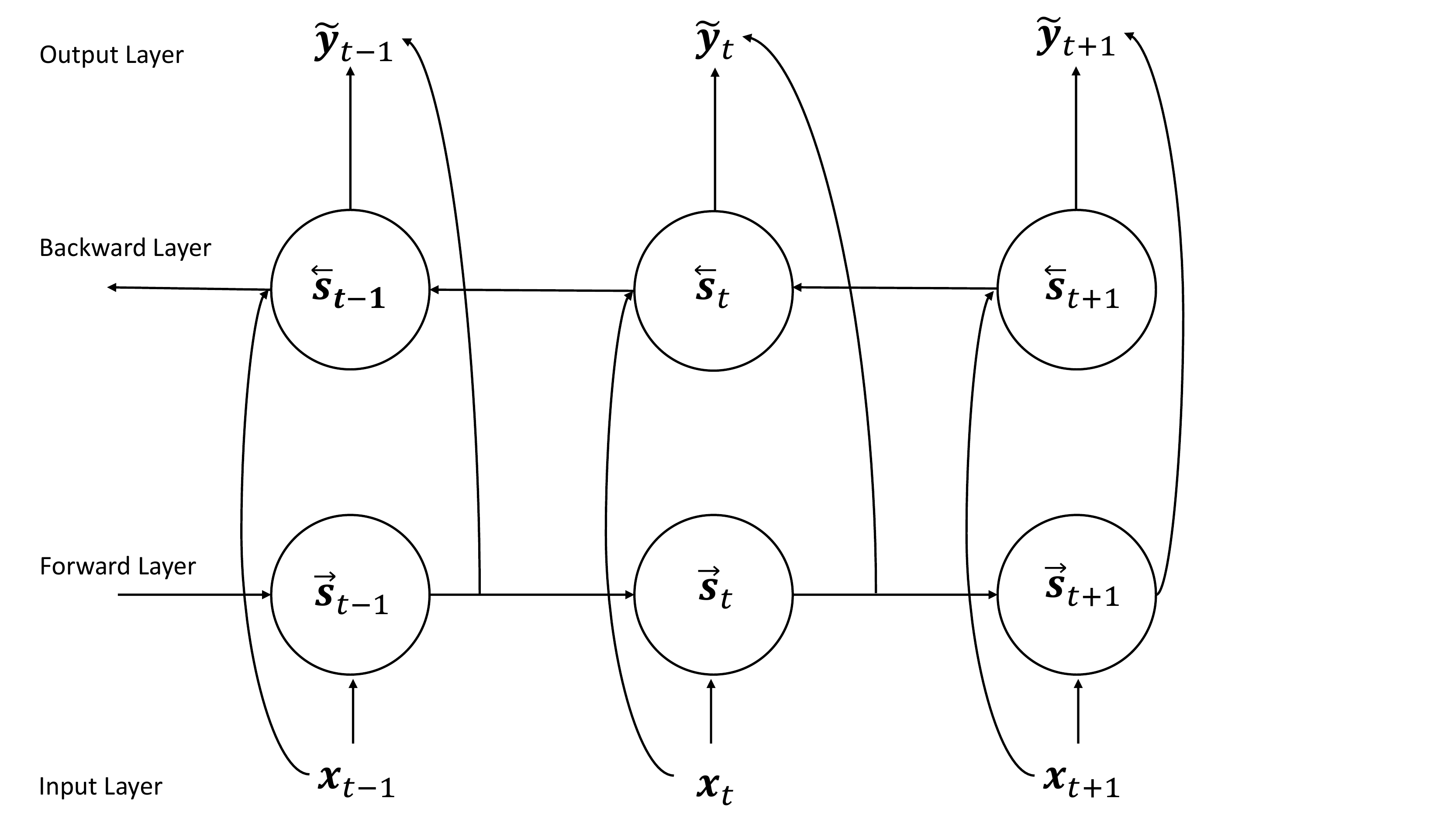}} \label{BiLSTM_a}
    \caption{Graphical illustration of time-series NNs, i.e., LSTM and BiLSTM. Fig.\,$1$\,(a) shows single-cell of an LSTM, where input is real-valued CSI realization for time instant $t$. Fig.\,$1$\,(b) depicts a fully connected BiLSTM-based NN, where inputs are learned in two ways. Circular shapes given in Fig.\,$1$\,(b) denote a single-cell of LSTM, which is given in Fig.\,$1$\,(a). The outputs of BiLSTM portray predicted CSI realizations.}
    \label{LSTM-BiLSTM}
\end{figure*}

\subsection{Bidirectional long-short term memory (BiLSTM)}\label{BiLSTM}
BiLSTM is composed of two independent long-short term memory (LSTM) networks. In a BiLSTM model, information is learned from both ends of the input data vector, which results in better prediction performance than traditional unidirectional LSTM. LSTM is an advanced version of RNN. RNN suffers from vanishing and exploding gradient problems in the backpropagation during model training. In the training phase, when the backpropagation algorithm advances backwards, i.e., from output to the input layer of RNN, the gradient often becomes smaller and smaller. By the time gradient reaches the input layer, it sometimes results in zero, not updating the weights of RNN; consequently, RNN cannot learn. Contrarily, exploding gradient turns to increase the gradient, which results in increasing weights of RNN; hence, gradient descent diverges. To this end, an LSTM-based NN was developed by Hochreiter and
Schmidhuber \cite{hochreiter1997long}. Schematic of an LSTM is depicted in Fig.\,\ref{LSTM-BiLSTM}. The core idea of LSTM is the introduction of a memory cell and multiplicative gates, which regulate the flow of information. Briefly, forget gate decides the amount of information to be stored in the cell by utilizing the current input, $\mathbf{x}_t$, and the output of the previous LSMT cell, denoted by $\mathbf{s}_{t-1}$. Mathematically,
    \begin{equation}
    \mathbf{f}_t= \sigma (\mathbf{W}_{f}\mathbf{x}_t+\mathbf{V}_{f}\mathbf{s}_{t-1}+\mathbf{b}_f)\\
    \label{forget}
\end{equation}
where
    \begin{equation}
    \sigma(x)= \frac{1}{1+e^{-x}} 
\end{equation}
is the \textit{sigmoid} activation function, $\mathbf{W}$ and $\mathbf{V}$ are the weight matrices, $\mathbf{b}$ is the bias vector, subscript $f$ is associated with the forget gate.  

The input gate determines the amount of information to be added into cell state $\mathbf{c}_{t-1}$ by exploiting $\mathbf{x}_t$ and $\mathbf{s}_{t-1}$. Mathematically,

     \begin{equation}
     \begin{aligned}
    \mathbf{i}_t &= \sigma (\mathbf{W}_{i}\mathbf{x}_t+\mathbf{V}_{i}\mathbf{s}_{t-1}+\mathbf{b}_i)\\
        \mathbf{g}_t &= \xi (\mathbf{W}_{g}\mathbf{x}_t+\mathbf{V}_{g}\mathbf{s}_{t-1}+\mathbf{b}_g)\\
        \label{input}
        \end{aligned}
\end{equation}   
where $\xi (\cdot)$ represents \textit{hyperbolic tangent} activation function, subscript $i$ and $g$ are associated with input gate. By utilizing input and forget gates,
LSTM can determine the amount of information to be retained and removed. Finally, the output gate calculates the output of LSTM cell by using an updated cell state, $\mathbf{c}_{t}$, and $\mathbf{x}_t$; the resultant output, given below, is then passed to the next LSTM cell of the network.  
    \begin{equation}
    \mathbf{o}_t= \sigma (\mathbf{W}_o\mathbf{x}_t+\mathbf{V}_{o}\mathbf{s}_{t-1}+\mathbf{b}_o)\\\:.
\end{equation}

As a result of the operations above, few information is dropped and a few is added, this updates the next long-term state as follows:
\begin{equation}
    \mathbf{c}_t= (\mathbf{f_t}\otimes \mathbf{c}_{t-1}+\mathbf{i}_t\otimes \mathbf{g}_t)\:
\end{equation}
where $\otimes$ represents Hadamard product. Lastly, short-term memory state, $\mathbf{s}_t$, is calculated by passing long-term memory, $\mathbf{c}_t$, through output gate as
\begin{equation}
    \mathbf{s}_t= \mathbf{o}_t\otimes \xi (\mathbf{c}_t)\:.
\end{equation}

In BiLSTM architecture, as depicted in Fig.\,\ref{LSTM-BiLSTM}, input information is learned in two directions, i.e., left-to-right (forward layer) and right-to-left (backward layer). 
Importantly, notation $t+1$ in BiLSTM architecture is only used for the illustration purpose, such indexes are based on passed CSI observations. The output information of each direction, denoted by $\vec{s}$ and $\cev{s}$, respectively, is passed simultaneously to the output layer, where output is calculated as
\begin{equation}
    \widetilde{\mathbf{y}}_t=\vec{\mathbf{s}}_t\otimes\cev{\mathbf{s}}_t\:.
\end{equation}
\subsection{Hybrid Model}\label{Hybrid}
\begin{figure}[t]
\centering
\includegraphics[scale=0.27,trim=1cm 6cm 1cm 3cm]{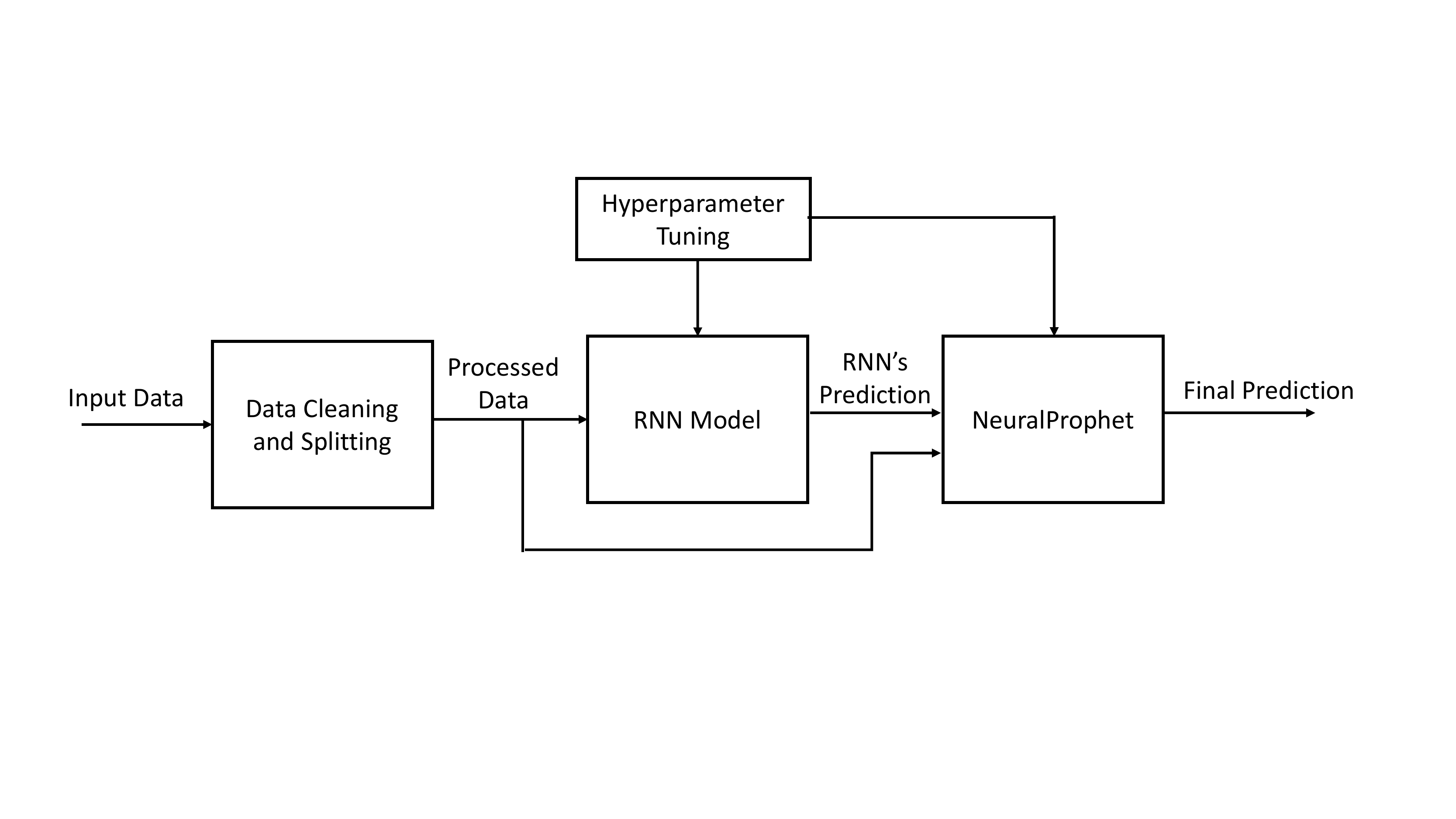}
\caption{Flow diagram of hybrid model.}
\label{hybrid_model}
\end{figure}

In the hybrid model, as shown in Fig.\,\ref{hybrid_model}, we utilize an RNN-based CP, summarized in Section\,\ref{RNN}, and NP, explained in the following subsection. In the beginning, dataset is cleaned, i.e., to check if there is corrupted/duplicate/missing data in the dataset. Additionally,  the data splitting is performed, in which dataset is divided into three sets i.e., training, validation, and test. Further, the input sequences are transformed into proper format, acceptable by ML models. Processed dataset is fed to the input of RNN, as well as NP. Then, we consider the predicted channel vector of RNN\footnote{It is, however, important to mention that the predicted output of BiLSTM can also be considered. But for the sake of lower computationally complexity of the hybrid model, we used RNN, as BiLSTM is computationally more expensive than RNN. The rationale behind this is two-way learning of BiLSTM and use of gates, e.g., input, forget.}, which is fed to NP along with input feature vector\footnote{In all CPs, input features, and corresponding labels are same.}. NP learns to correct the predicted output of RNN. Later in Section\,\ref{results}, we demonstrate that NP can significantly correct the predicted outputs of RNN, and it outperforms all standalone models, notably BiLSTM. In the following, we explain working functionality of NP. 
\subsubsection*{NeuralProphet (NP)}\label{NP}
Within a short span of time, NP has emerged as a promising choice for different time-series prediction tasks \cite{neuralprophet}. 
NP is an explainable, and scalable prediction framework.
It is sometimes important to analyze the performance of a prediction model in the form of different components. DL-based models are difficult to interpret due to their black-box nature. Contrarily, NP  is composed of different components, where each component contributes to predicted output, and their behaviour is well interpretable.

NP is composed of various components\footnote{In the documentation of NP, it is composed of six components. However, with regard to our application of channel prediction, we dropped a few as some of them are irrelevant for CSI prediction, e.g., holidays. For more details, interested reader can refer to \cite{neuralprophet}.}, which contribute additively to the prediction. Each component is composed of individual inputs and modelling methodology. The output of each component is $D$-step ahead future CSI realizations. For the notational convenience, we will explain the model with $D=1$, which will be later extended for multi-step, i.e., $D$-future steps. In the context of channel prediction, the predicted value of NP for time instant $t$ can be written as \cite{neuralprophet}

\begin{equation}
    \widetilde{z}_t= R_t+ F_t+ A_t
\end{equation}
where $R_t$ and $F_t$ represent trend and seasonality functions for input data, respectively. Also, $A_t$ is the AR effect for time $t$ based on previous CSI realizations. Below, we explain each of them. 

\subsubsection*{$R_t$}The trend function concerns the overall variation in the input data. It tries to learn the points where clear variation in the data occurs; these points are called change-points, represented by $\{n_1, n_2, \dots, {n_m}\}$, composed of a total of $m$ change-points (tuned using \textit{grid search}). A trend function can be expressed as
\begin{equation}
    R_t= (\zeta^0+(\mathbf{\Gamma}_t)^{\dag})\boldsymbol{\zeta})\cdot t+(\rho^0+(\mathbf{\Gamma}_t)^\dag\boldsymbol{\rho})
\end{equation}
where $\boldsymbol{\zeta}=\{\zeta^1, \zeta^2, \dots, \zeta^{m}\}$, and $\boldsymbol{\rho}=\{\rho^1, \rho^2, \dots, \rho^{m}\}$, are the vectors of growth rate and offset adjustments, respectively, and $\boldsymbol{\zeta}\in\mathbb{R}^{m}$ and $\boldsymbol{\rho}\in\mathbb{R}^{m}$. Besides, $\zeta^0$ and $\rho^0$ are the initial growth rate and offset values, respectively. And, $\boldsymbol{\Gamma}_{t}=\{\Gamma^{1}_{t}, \Gamma^{2}_{t}, \dots, \Gamma^{m}_{t}\}$, where $\boldsymbol{\Gamma}_t\in\mathbb{R}^{m}$, which represents whether a time $t$ has past each change-point. Also, $(\cdot)^{\dag}$ representing transpose. For a $j^{th}$ change-point, $\Gamma^j_t$ is defined as

\begin{equation}
    \Gamma^j_t=
    \begin{cases}
      1, & \text{if}\ t\geq n_j \\
      0, & \text{otherwise}
    \end{cases}\:.
  \end{equation}
\subsubsection*{$F_t$} The seasonality function, modeled using Fourier terms \cite{Fourier_Time_Series}, captures periodicity in the dataset, and is expressed as
\begin{equation}
    F^{p}_t=  \sum_{r=1}^{k} \bigg(a_r\cdot cos\bigg(\frac{2\pi rt}{p}\bigg)+b_r\cdot sin\bigg(\frac{2\pi rt}{p}\bigg)\bigg)
\end{equation}
where $k$ is the number of Fourier terms, which are defined for a seasonality having periodicity $p$. At a time step $t$, the effect of all seasonalities can be expressed as
\begin{equation}
    F_t= \sum_{p\in \mathbb{P}}F^p_t
\end{equation}
where $\mathbb{P}$ is the set of periodicities. We considered three seasonalities, their values are written in Table\,\ref{table:I}.  
\subsubsection*{$A_t$} AR is the process of regressing a CSI future realization against its past realizations. The total number of past CSI realizations considered in the AR process are referred to as the order, denoted as $d$, of AR process. A classic AR process of order $d$ can be modeled as
\begin{equation}
    z_t= q+ \sum_{e=1}^{e=d}\theta_e\cdot z_{t-e}+\epsilon_t
\end{equation}
where $q$ and $\epsilon_t$ are the intercept and white noise, respectively, and $\theta$ are the coefficients of AR process. 
The classic AR model can only make one-step ahead prediction, and to make multi-step ahead prediction, $D$ distinct AR models are required to fit. To this end, we utilize feed-forward NN along with AR (built-in feature of NP), termed as \textit{AR-Net} \cite{AR-Net}, to model AR process dynamics. NP-based \textit{AR-Net} can produce multi-step future CSI realizations by using one AR model. \textit{AR-Net} mimics a classic AR model, with the only difference of data fitting. \textit{AR-Net} is a feed-forward NN that maps AR model.    
 
In the \textit{AR-Net}, $d$ last observations of CSI realizations are given as input, denoted as $\mathbf{z}$, which are processed by the first layer and then passed through each hidden layer. Correspondingly, $D$-step ahead future CSI realizations, denoted by $\widetilde{\mathbf{z}}=\{A_t, A_{t+1}, \dots, A_{t+D}\}$, can be obtained at the output layer. Mathematically, 

\begin{equation*}
\begin{aligned}
        \boldsymbol{\omega}^\text{out}_1&=\alpha(\mathbf{U}_1\mathbf{z}+\mathbf{b}^\text{NP}_1)\\
        \boldsymbol{\omega}^\text{out}_i&=\alpha(\mathbf{U}_{i}\boldsymbol{\omega}^\text{out}_{i-1}+\mathbf{b}^\text{NP}_i)\quad\quad \text{for}\quad i\in[2, 3, \dots, l]\\
        \mathbf{\widetilde{z}}&=\mathbf{U}_{l+1}\boldsymbol{\omega}^\text{out}_l
\end{aligned}
\end{equation*}
where $\alpha(\cdot)$ is the \textit{rectified linear unit (ReLu)} activation function, written as
\begin{equation}
    \alpha(\gamma)=
    \begin{cases}
      \gamma, & \gamma\geq 0 \\
      0, & \gamma<0
    \end{cases}\:.
  \end{equation}
Further, $l$ is the number of hidden layers having dimension of size $n_h$, $\mathbf{b}^{\text{NP}}\in\mathbb{R}^{n_h}$ is the vector of biases, $\mathbf{U}\in\mathbb{R}^{n_h\times n_h}$ is the weight matrix for hidden layers, except for the first $\mathbf{U}_1\in\mathbb{R}^{n_h\times d}$ and last $\mathbf{U}_{l+1}\in\mathbb{R}^{D\times n_h}$ layers. 
In the AR component of NP, an important selection parameter is the order of AR, i.e., $d$, which is hard to select in practice. In general, $d$ is chosen such that $d=2D$.
\begin{figure*}[t]
\begin{center}
  \includegraphics[width=18cm, height=5cm]{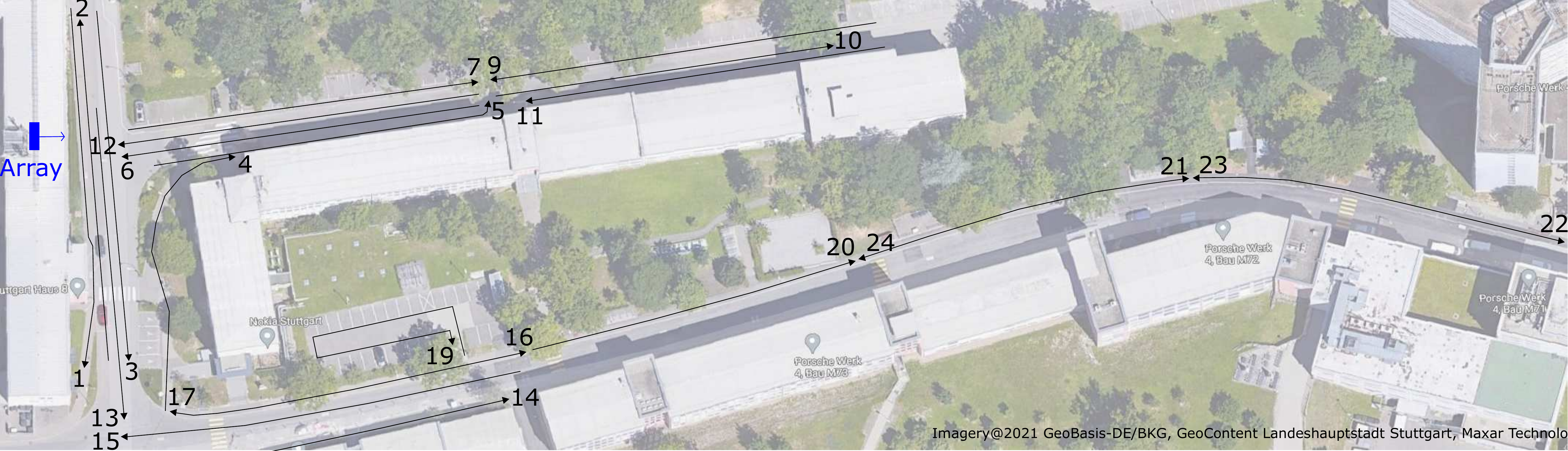}
\caption{{Pictorial representation of measurement tracks in the Nokia campus. The blue bar on the extreme left side of the figure shows a mMIMO antenna located on the rooftop, transmitting in the direction marked with the blue arrow. Tracks on which the UE was moving are drawn with black lines, where the arrowhead shows the direction of the UE's movement. Furthermore, numerical values depict the track numbers. Source: \cite{Karam_WCL}.}}\label{Map}      
\end{center}
\end{figure*}

\section{Real-Time Channel Measurements}\label{Measurement Campaign}

We utilize a dataset, which was recorded at Nokia Bell-Labs. 
Briefly, a vehicle (mimicking a UE) was considered, moving on the sketched tracks, as indicated in Fig.\,\ref{Map}. A transmitter is equipped with $64$ antennas placed on a rooftop (height of approximately $15\,$meters), while UE was equipped with a single monopole antenna mounted at the height of $1.5\,$meters on the receiver cart, i.e., UE. The UE was moving at a speed of $3$ to $5\,$kmph, and array antennas transmitted $64$ time-frequency orthogonal pilot signals at $2.18$\,GHz carrier frequency. For more details of the measurement campaign, interested reader can refer to \cite{Karam_WCL}.  

\section{Experimental Setup and Results}\label{results}
In this section, we analyze the prediction capability of the hybrid model and compare its performance with other CPs. 
To train CPs, we used dataset of track-1, which is composed of $116$k consecutive CSI realizations, i.e., $\{\mathbf{h}_t|t=1, \dots, 116\text{k}\}$, where interval of each recorded value is $0.5\,$ms. The dataset (normalized) is passed through necessary pre-processing and formatting steps using custom-built input pipelines so that it can easily parsed through each model. 
Further, the dataset of track-1 is divided into three parts: training (80\%), validation (10\%), and test (10\%). For the sake of simplicity, the dataset of the first four transmit antennas is extracted for performance evaluation. Simulation parameters are given in Table\,\ref{table:I} unless stated otherwise, and parameter details are given in Section\,\ref{NP} and \ref{hyper}. The training process starts from an initial state, where weights and biases are randomly initialized. At the $t^{th}$ time iteration, pre-processed\footnote{Real and imaginary parts of channel are separated, delayed channel realizations and corresponding labels are created, etc.} channel vector $\mathbf{h}_t$ is fed as an input to CPs. Recalling  Fig.\,\ref{LSTM-BiLSTM}, $\mathbf{x}_t=\mathbf{h}_t$ and $\widetilde{\mathbf{y}}_t=\widetilde{\mathbf{h}}_{t+D}$. For notational convenience, let us assume that input for each CP and corresponding $D$-step ahead prediction is denoted by $\mathbf{x}^{\tau}$ and $\widetilde{\mathbf{y}}^{\tau}$, respectively. 
\begin{table} {}
\caption{Simulation Parameters}
\centering
 \begin{tabular}{|c| c| c| c|}
 \hline
 \textbf{Parameter} & \textbf{Value} & \textbf{Parameter} & \textbf{Value} \\ 
 \hline
 $\{l, \beta\}$ & $\{3, 1\}$ & $\{\gimel, n_h\}  \text{(NP)}$ & $\{0.01, 32\}$ \\ [1ex]
 \hline
 $\text{Epochs}$ & $50$ & $\{d,D\}$ & $\{48,24\}$ \\

 \hline
 $m$ & $30$ & $\{\gimel, n_h\}\text{(RNN\:\&\:BiLSTM)}$ & $0.001, 200$  \\
\hline
 \multirow{1}{*}{{$\{(k, p)\}$}}  &
      \multicolumn{3}{c|}{$\{(6, 365.25), (3, 7), (6, 1)\}$}  \\
 \hline
\end{tabular}

\label{table:I}
\end{table}
To train CPs, we used Huber loss as a cost function, which is defined as
\begin{equation}
    L_{\text{huber}}(\mathbf{y}^{\tau},\widetilde{\mathbf{y}}^{\tau})=
    \begin{cases}
      \frac{1}{2\beta}(\mathbf{y}^{\tau}-\widetilde{\mathbf{y}}^{\tau})^2, & \text{for}\ |\mathbf{y}^{\tau}-\widetilde{\mathbf{y}}^{\tau}|\leq\beta \\
      |\mathbf{y}^{\tau}-\widetilde{\mathbf{y}}^{\tau}|-\frac{\beta}{2}, & \text{otherwise}
    \end{cases}\:
  \end{equation}
where ${\mathbf{y}^{\tau}}=\{{y}^{\tau}_{t+1}, {y}^{\tau}_{t+2}, \dots, {y}^{\tau}_{t+D}\}$ are corresponding labels for the input CSI realizations $\mathbf{x}^{\tau}=\{{x}^{\tau}_{t-1},{x}^{\tau}_{t-2},\dots,{x}^{\tau}_{t-d}\}$, and $\widetilde{\mathbf{y}}^{\tau}= \{
{\widetilde{y}}^{\tau}_{t+1}, \widetilde{y}^{\tau}_{t+2}, \dots, \widetilde{y}^{\tau}_{t+D}\}$ is the $D$-step ahead predicted channel vector. 
By using Huber loss as a cost function, a batch of $32$ samples is fed into CPs, the predicted outcome is compared with true labels, and error is backpropagated to update weights and biases using \textit{adaptive moment estimation} (\textit{Adam}) as an optimizer.
The training iterations are repeated until a certain convergence condition is satisfied, i.e., cost function is below a threshold value.  
Furthermore, open-source libraries such as \textit{TensorFlow}, \textit{Keras}, \textit{Scikit-learn}, and \textit{Pandas}, are used for implementation of CPs. Once the CPs are trained, they can produce predictions in large batches of data on desired future steps, i.e., $D$.

\subsection{Hyperparameter Tuning}\label{hyper}
Selecting the best training parameters, e.g., learning rate ($\gimel$), number of hidden layers ($l$) and units ($n_h$), for computational models is critical for achieving better prediction accuracy. 
As hyperparameters are not directly learned by a model, rather, we manually define them before fitting the model. Therefore, a well-known automated strategy, i.e.,  \textit{grid search}, is used for hyperparameter tuning. 
For NP, \textit{discontinuous growth} function and $90\%$ \textit{change-point range} gave best results, other tuned parameters for NP and other CPs are listed in Table\,\ref{table:I}. Also, the standard NP model only takes uni-variate data to produce its output. In our hybrid model, we adapted the standard NP model for multivariate data by adding extra future regressor into the NP model. Future regressor are the variables, which are known into the future. In our case, predictions of RNN are known to us and we included this information as a future regressor of NP model. In this way, we feed both RNN predictions and true sequences as multivariate input to NP model to enhance its prediction capability. 
Besides, in RNN and BiLSTM CPs, the dropout layer is adapted, which drops hidden units randomly with a probability of $0.2$, to prevent over-fitting. 

\subsection{Evaluation Parameters}
The performance of four CPs, i.e., NP, RNN, BiLSTM, and hybrid model, is evaluated by using normalized mean-squared error (NMSE) and cosine similarity. Suppose that multi-step ahead channel predicted vector and corresponding true labels are represented by $\widetilde{\mathbf{h}}_{t+D}$ and $\mathbf{h}_{t+D}$, respectively. Then NMSE, for a single track, is defined as
\begin{equation}
    {\text{NMSE}}=\mathsf{E}\left\{\frac{{} 	\left\|{{{\widetilde{{\mathbf{h}}}_{t+D}}}-{{\mathbf{h}}_{t+D}}}\right\|^2_{2}}{{	\left\|{{{{\mathbf{h}}_{t+D}}}}\right\|^2_{2}}}\right\}\:
    \label{NMSE}
\end{equation}
where $\mathsf{E}\{\cdot\}$ denotes expectation operator, and $\|\cdot\|^2_{2}$ represents squared Euclidean norm. The second evaluation parameter, i.e., cosine similarity, is expressed as
\begin{equation}
    \text{Cosine Similarity}=\mathsf{E}\left\{\frac{\mid\widetilde{\mathbf{h}}_{t+D}^{*}\ \mathbf{h}_{t+D}\mid }{{\left\|{({{{\widetilde{\mathbf{h}}_{t+D}}}})}\right\|_{2}}  {\|{{{\mathbf{h}}_{t+D}}}\|_{2}}}\right\}\:
\end{equation}
where $|\cdot|$ and $(\cdot)^{*}$ denote absolute value and conjugate transpose, respectively. 
\subsection{Performance Comparison}
\begin{figure*}[ht]
    \centering
    \subfigure[NP]{\includegraphics[width=0.24\textwidth]{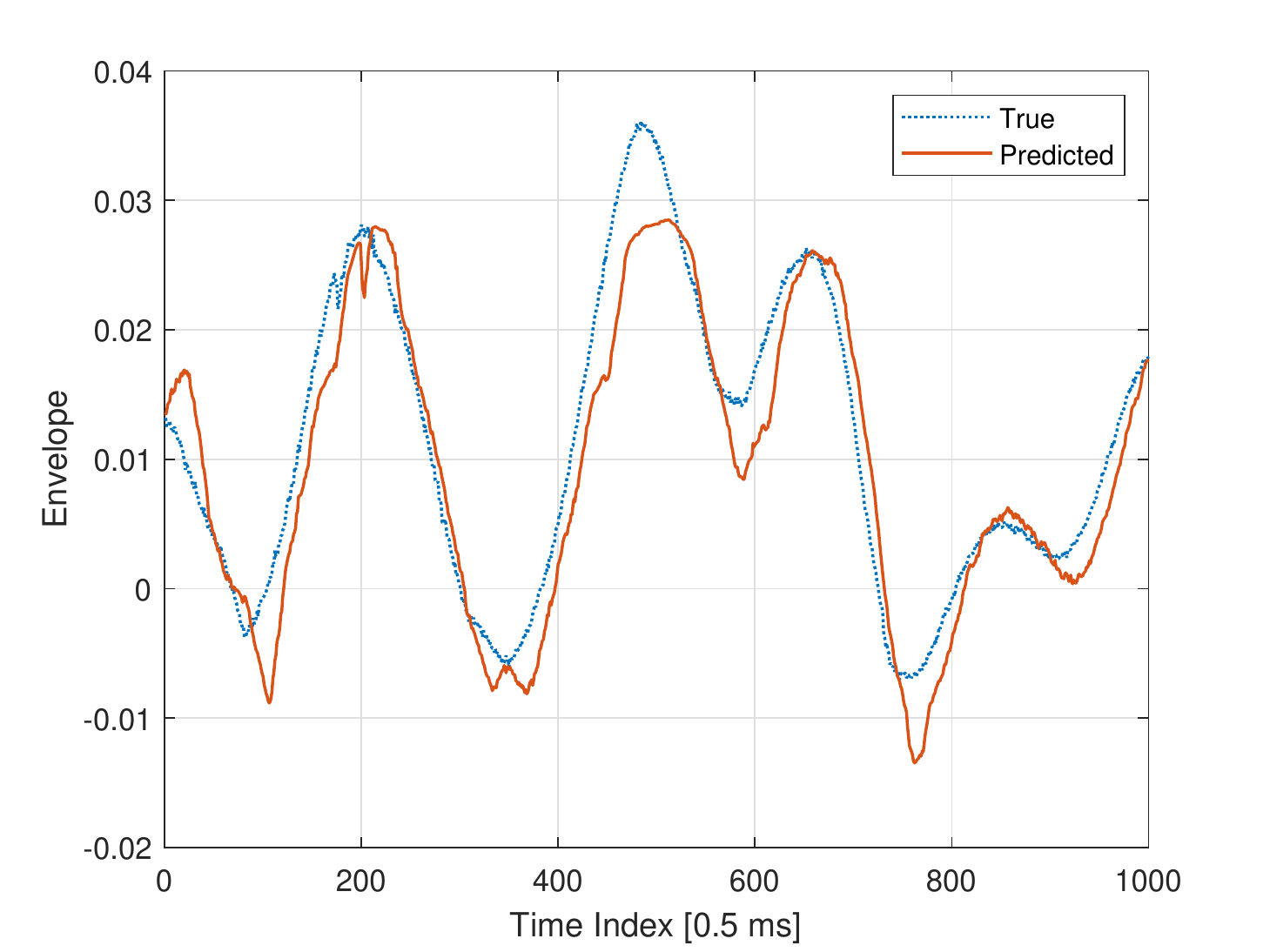}} 
    \subfigure[RNN]{\includegraphics[width=0.24\textwidth]{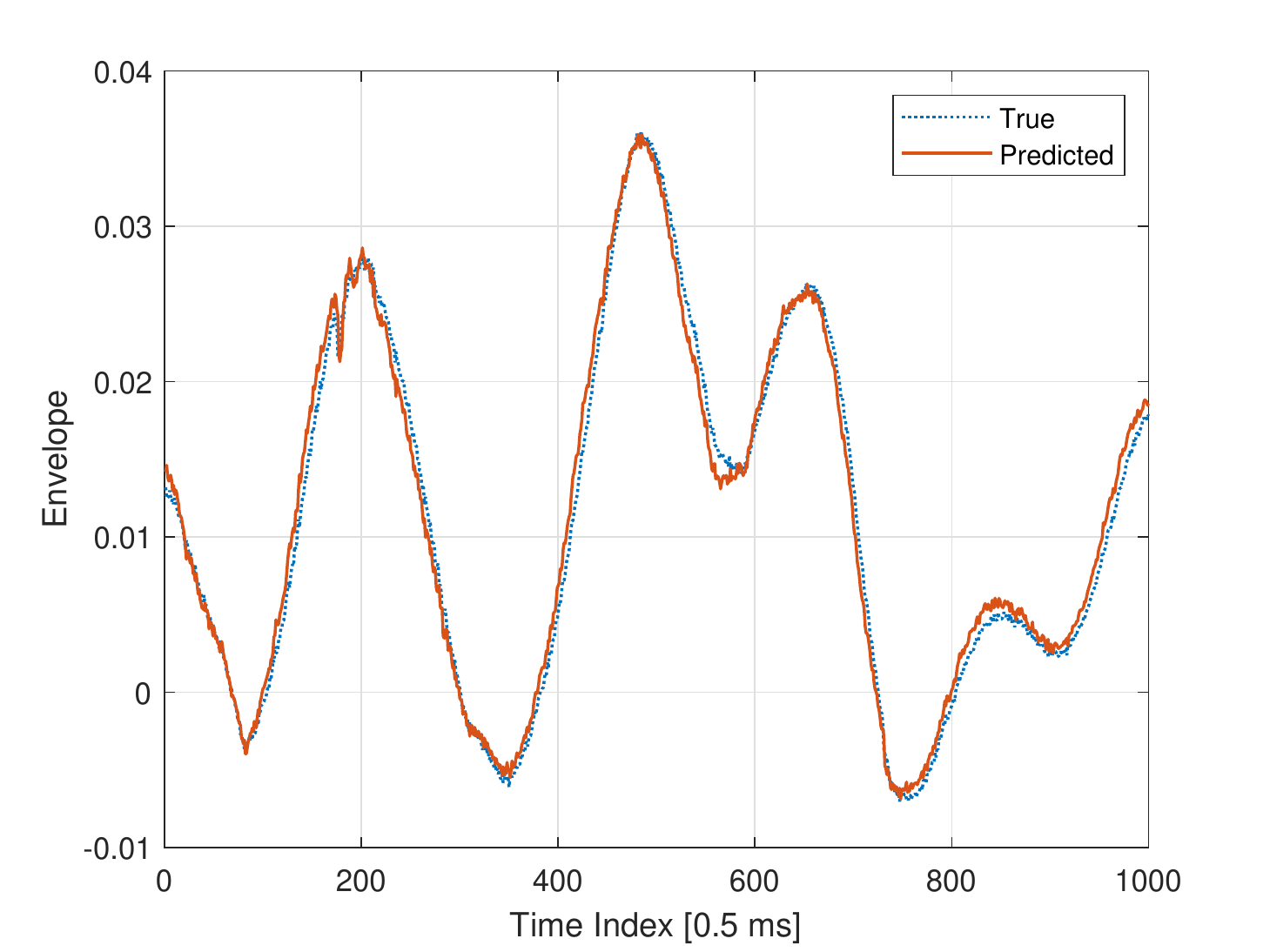}} 
    \subfigure[BiLSTM]{\includegraphics[width=0.24\textwidth]{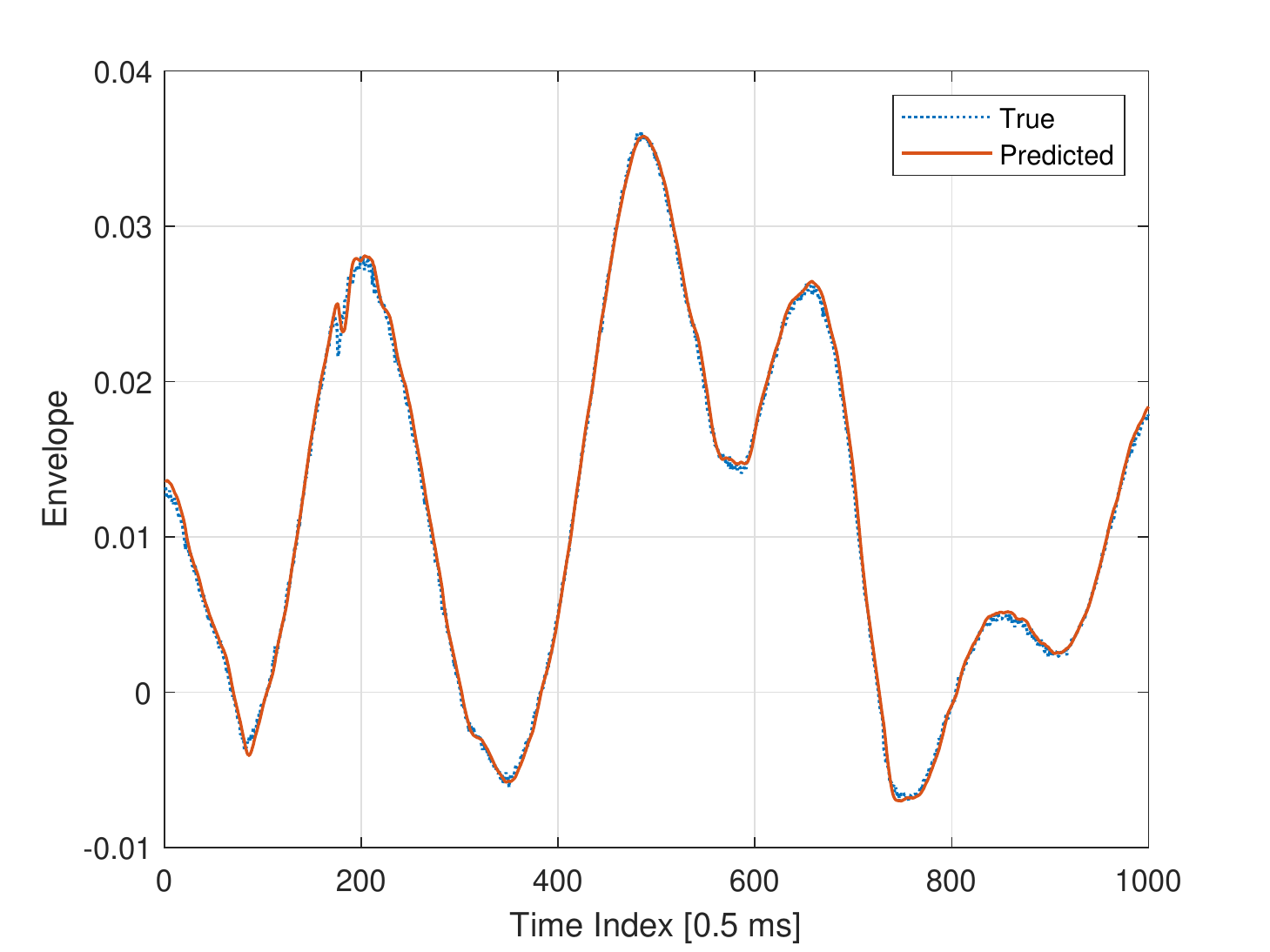}}
        \subfigure[Hybrid]{\includegraphics[width=0.24\textwidth]{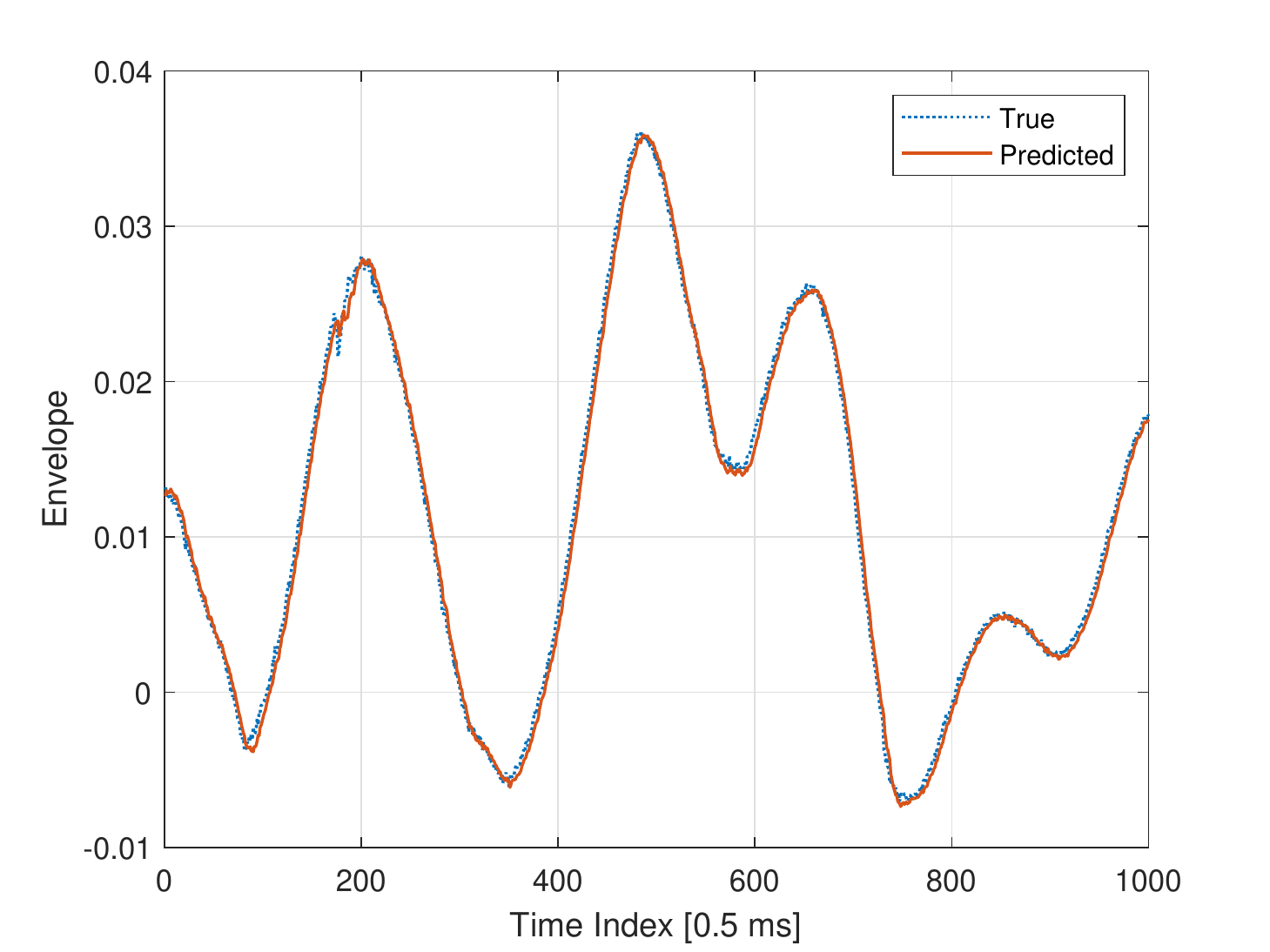}}
    \caption{A snapshot of predicted and true channel, where imaginary part of transmit antenna number three is plotted. The UE was moving on track-1, and each measurement is captured every 0.5\,ms. }
    \label{fig_visual}
\end{figure*}
A visual illustration of CPs is shown in Fig.\,\ref{fig_visual}, where CSI realizations are shown for both predicted and the true CSI. The internal configuration and layers structure of RNN and BiLSTM are kept similar for an impartial comparison. It can be observed that NP's performance is lower, which is due to the reason for using a simple feed-forward NN. RNN has a slightly better performance in comparison to NP. However, BiLSTM and hybrid model give most accurate results.
In the following, we explain results, which are averaged over the entire test dataset of each track to show a clear winner.

\begin{figure}[t]
\centering
\includegraphics[scale=0.59]{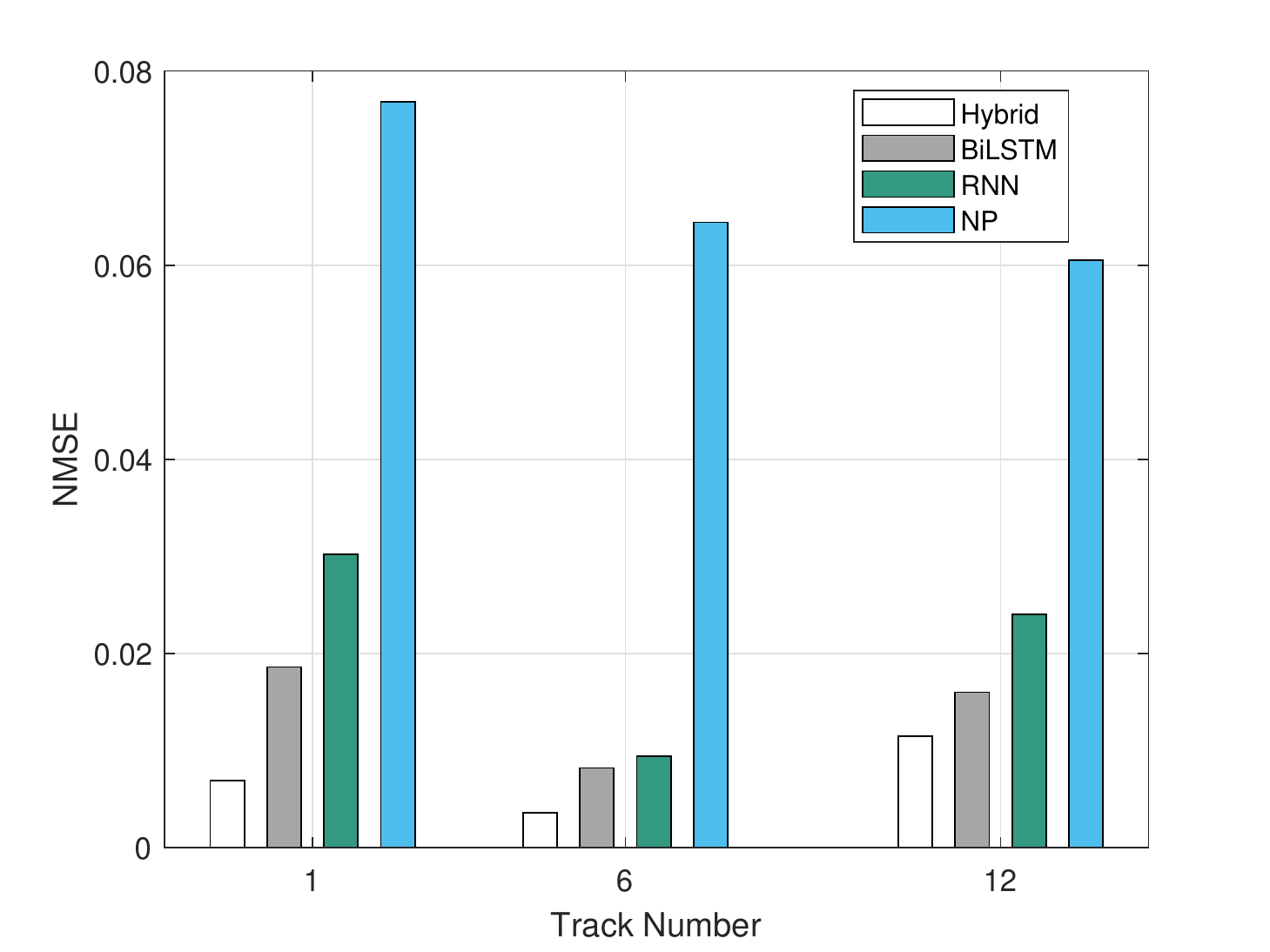}
\caption{Comparison of different models on three different tracks, where NMSE results are independent with respect to track number as each track has different channel strength.}
\label{NMSE_G1}
\end{figure}

Fig.\,\ref{NMSE_G1} shows the comparison of four CPs in terms of NMSE. Particularly, performance is evaluated on three different tracks, which were unseen by the CPs during the training phase. The trend reveals that NP, when used standalone, does not perform well; hence, it gives the worst performance. On the other hand, BiLSTM outperforms RNN, which is due to the fact that BiLSTM can retain information in their memory for longer time periods and learns the input data in both directions, thereby performing better. In contrast, the hybrid model performs better than other CPs. The rationale behind this is that NP learns statistical information of RNN's predicted values and true inputs and compares them with the corresponding true labels. Hence, it improves the performance of RNN's predictions. For instance, on track-1, there is approximately $4\,$dB NMSE reduction in the hybrid model as compared to BiLSTM.    

\begin{table}{} 
\centering
\caption{Performance comparison on different tracks and CPs}
\begin{tabular}{ c c|c c} 
\hline
Track Number & Model & NMSE &$\text{Cosine Similarity}$ \\
\hline

\multirow{4}{*}{1} & Hybrid& $\mathbf{0.006}$&$\mathbf{0.997}$\\ 

&BiLSTM&	$0.018$&$0.995$\\
&RNN&	$0.030$&$0.989$\\
&NP&	$0.076$&$0.978$\\

\hline
\multirow{4}{*}{6} & Hybrid& $\mathbf{0.003}$&$\mathbf{0.999}$\\ 

&BiLSTM&	$0.008$&$0.998$\\
&RNN&	$0.009$&$0.996$\\
&NP&	$0.064$&$0.986$\\

\hline
\multirow{4}{*}{12} & Hybrid& $\mathbf{0.011}$&$\mathbf{0.995}$\\ 

&BiLSTM&	$0.016$&$0.994$\\
&RNN&	$0.024$&$0.991$\\
&NP&	$0.060$&$0.989$\\

\hline



\hline
\end{tabular}
\label{table:2}
\end{table}

In Table\,\ref{table:2},
the comparison shows that the proposed hybrid model outperforms BiLSTM, RNN, and NP. For instance, on track-12 that is significantly away from trained track, hybrid model gives NMSE of $11\times10^{-3}$ whereas NP has $60\times10^{-3}$. Also, hybrid model brings significant gain in comparison to DL models, i.e., RNN and BiLSTM. Similarly, hybrid model improves cosine similarity.   
\section{Conclusion} \label{conclusion}
In this paper, we have developed various time-series models for mMIMO channel prediction. In particular, we evaluated the performance of RNN, BiLSTM, NP, and proposed hybrid framework comprising RNN and NP. We learned from numerical results, which are obtained by exploiting real-time channel measurements, that the proposed hybrid framework outperforms standalone RNN, and NP. Furthermore, results demonstrate the ability of hybrid framework to overcome the performance of advanced DL model, i.e., BiLSTM. 

In the future, we will evaluate the performance, both from gain and complexity perspective, of the above models and others, for instance, Kalman filter. Furthermore, we aim to explore NP with different NN architectures, e.g., replacing feed-forward NN with BiLSTM.

\bibliographystyle{IEEEtran}
\bibliography{main}
\end{document}